\documentclass[aps,pra,reprint,showpacs,superscriptaddress,twocolumn,10pt]{revtex4-1}%
\usepackage{amsfonts}
\usepackage{mathrsfs}
\usepackage{amsmath}
\usepackage{amssymb}
\usepackage{graphicx}
\usepackage{color}%
\usepackage{mathtools}
\usepackage{booktabs}
\usepackage{mleftright}
\usepackage{float}
\usepackage[colorlinks,linkcolor=blue,citecolor=blue,urlcolor=blue,hyperindex,bookmarks=false,pdfstartview=FitH]{hyperref}
\usepackage[utf8]{inputenc}
\date{\today}
\begin{document}
\bibliographystyle{plainnat}

\title{Manifestation of topological phase in neutron spin rotation without adiabatic regime}

\author{Jian-Jian Cheng}
\email{chengjianjian@csrc.ac.cn}
\affiliation{Beijing Computational Science Research Center, Beijing 100193, China}

\begin{abstract}
The Bitter-Dubbers (BD) experiment is an important experiment that originally aimed to measure  topological phase using polarized-neutron spin rotation in a helical magnetic field under adiabatic conditions. Contrary to expectations, upon reevaluation of the BD experiment, it has been found that adiabatic conditions are not  necessary for measuring topological phase. In scenarios where the magnetic field is neither homogeneous nor strong enough, and the neutron has a fast velocity, the topological phase can still be manifested.
To demonstrate this, we analytically solve the time-dependent Schr$\ddot{\textrm{o}}$dinger equation for the neutron spin rotation in general rotating systems. These exact solutions are then utilized to investigate the nonadiabatic  topological phase under the  conditions mentioned above. The numerical simulations of the nonadiabatic  topological phase have shown a strong concurrence with the BD experimental data.  This novel result extends our understanding of the  topological phase observed in neutron spin rotation, even in more complex and dynamic scenarios beyond the originally required adiabatic conditions.

\end{abstract}

\maketitle

\textit{Introduction}.\textemdash The study of quantal phase factors, both experimentally and theoretically, has always been a fascinating topic, captivating researchers with its complexities and revealing unexpected features~\cite{phase,geo}.
In 1984, Michael Berry wrote a famous paper discussing the adiabatic evolution of eigenenergy states when the external parameters of a quantum system change slowly, forming a loop in the parameter space. In the absence of degeneracy, the eigenstate will return to itself after completing the loop, but during the adiabatic evolution, the state may acquire both a dynamical phase and an additional geometric phase  (also termed as the Berry phase)~\cite{Berry}.
Following Berry's work, numerous experimental demonstrations of geometric phase have been conducted using various physical systems such as polarized light~\cite{light,light2}, nuclear quadrupole resonance~\cite{resonance}, neutron interferometers~\cite{BD,BD2,BD3,BD4,BD5,BD6}, and others~\cite{other1,other2}.

A typical example used to understand the concept of geometric phase involves a spin-1/2 particle subjected to a slowly varying magnetic field, which follows a closed path in the parameter space. This measurement of the geometric phase in a spin-1/2 particle system was observed in the Bitter-Dubbers (BD) experiment~\cite{BD}. In this experiment, the interaction between the neutron's spin and the helical magnetic field made the neutron's spin direction  rotate. By quantifying the polarization of the neutrons, valuable physical information regarding the behavior of neutrons in the magnetic field could be obtained.
 Berry's law proves that the spin state acquires an additional topological phase when the spin vector returns to its initial direction
under the adiabatic conditions. The manifestation of the additional topological phase does not depend on the interior dynamics of the system, but instead relies on its geometric history. This implies that the additional topological phase can be expressed as a line integral over the loop in the parameter space and is independent of the exact rate of change along the loop. The adiabatic conditions require the helical magnetic field to be homogeneous and strong enough, and the velocity of the neutron along the z-axis to be small enough~\cite{BD,Sun}. It is believed that the adiabatic approximations are necessary for the spin state to acquire the topological phase in the BD experiment. However, it is important to note that the adiabatic approximation is not always valid, thus  these experiments require very precise control.

When reexamining the BD experiment, it is essential to directly consider the scenarios where the magnetic field is neither homogeneous nor strong enough, and the neutron has a fast velocity. Instead of firstly studying the evolution of the spin state under adiabatic conditions and later extending it to nonadiabatic situations, it is worth exploring the effects that may arise under nonadiabatic conditions. This approach allows us to gain deep insight into how the system behaves without the adiabatic approximation.

Obtaining the analytical solutions of a particle in general rotating systems are challenging, but we can draw inspiration from the transitionless quantum driving method~\cite{STA}.
Based on this, we reevaluated the rotation of the neutron spin without adiabatic regime. Fortunately, we obtain the corresponding exact solution. Then the obtained exact solutions are utilized to investigate the nonadiabatic topological phase~\cite{jingque1,jingque2}. It is  found that the measured  topological phase data of neutrons reaching a velocity of $v=500\,\textrm{m/s}$ from the BD experiment can better match our current predictions of the nonadiabatic topological phase. Particularly, the first few data points, where the adiabatic conditions deviate more, fall below the predicted curve of the Berry phase.

\textit{Model}.\textemdash In 1987, T. Bitter and D. Dubbers conducted measurement of  topological phase using polarized-neutron spin rotation in a helical magnetic field. The BD experiment used a slow, monochromatic beam of polarized neutrons obtained from a neutron guide. This neutron beam enters and exits a field-free region within a Mumetal cylinder. The Mumetal cylinder is aligned with the beam in a coaxial manner. Inside the Mumetal cylinder, the neutrons pass through a static helical magnetic field. This magnetic field is perpendicular to the beam and completes one full right turn around the beam axis as the neutrons enter and exit (for more details see Ref.~\cite{BD}).
\begin{figure}[H]
\begin{center}
\includegraphics[width=180pt]{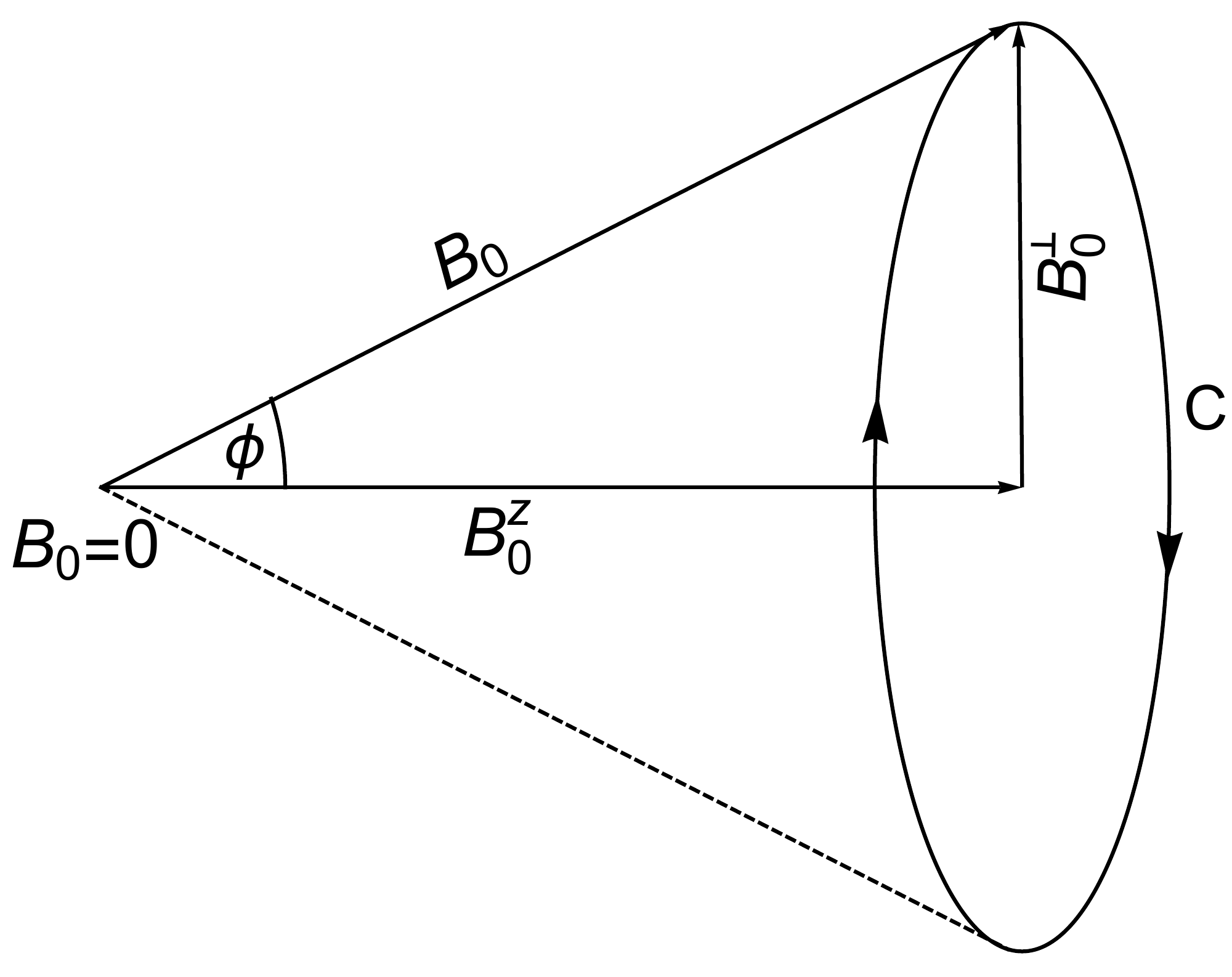}
\end{center}
\caption{The magnetic field vector $\textbf{B}_{0}(t)$ is around a closed loop $\textbf{C}$. The velocity of
the neutron beam is along the z-axis.}
\label{fig1}
\end{figure}
The neutrons on their right along the axis of the cylinder see a
right circularly polarized magnetic field $B_{0}^{\perp}$
 rotating by $2\pi$ over a length $L$, as illustrated in Fig.~\ref{fig1}. Therefore,
the neutrons perceive that the external magnetic field $\textbf{B}_{0}(t)$ varies over time, as
\begin{eqnarray}
\label{mod}
\begin{aligned}
\textbf{B}_{0}(t)=B_{0}[&\sin\phi(\cos\frac{2\pi v t}{L} \textbf{e}_{x}+\sin\frac{2\pi v t}{L}\textbf{e}_{y})\\
&+\cos\phi\,\textbf{e}_{z}].
\end{aligned}
\end{eqnarray}
Correspondingly, the Hamiltonian of the neutron in the varying magnetic field can be described as $(\hbar=1)$ 
\begin{eqnarray}
\label{cichang}
\hat{H}_{0}(t)=-\kappa\textbf{B}_{0}(t)\cdot\hat{S},
\end{eqnarray}
where $\kappa$ represents the magnitude of the gyromagnetic ratio for neutrons and $\hat{S}$
is the spin operator involving the Pauli matrices $\hat{S}=\frac{1}{2}\hat{\sigma}$.
In previous discussions of the BD experiment, the topological phase observed in neutron spin rotation was referred to the Berry phase~\cite{BD}. This phase emerges under adiabatic conditions where $\textbf{B}_{0}(t)$ is homogeneous and strong enough (for large $\textrm{L}$ and $\textrm{B}_{0}$), and the velocity $v$ of the neutron along the $z$ axis is small enough~\cite{Sun}.

\textit{Genuine dynamical evolution}.\textemdash It should be noted that achieving adiabatic conditions is a challenging task in real-world scenarios. When the magnetic field is neither homogeneous nor strong enough, and the neutron has a fast velocity along the $z$ axis, the neutrons, as they fly, encounter a rapidly varying magnetic field, leading to a time-dependent nonadiabatic Hamiltonian.
To investigate BD experiments under such nonadiabatic conditions, Sun \emph{et al}. proposed the higher-order adiabatic method~\cite{Sun} to deal with this system. Although this method is theoretically applicable to nonadiabatic parameter systems, it involves a considerable number of approximations, making analytical analysis challenging. Moreover, when the velocity of neutrons increases further, the higher-order adiabatic method may fail to be effective.

An alternative and successful approach proposed in this paper is inspired by the transitionless quantum driving method~\cite{STA}.
The transitionless quantum driving method provides a way to construct Hamiltonian for which the adiabatic approximation of the time-dependent wave function, evolving with a reference Hamiltonian, becomes exact by adding auxiliary interactions.
The physical realizability of the auxiliary Hamiltonian requires individual examination in each system. For example, if the reference Hamiltonian describes a particle in a time-dependent harmonic potential, the corresponding auxiliary Hamiltonian would involve nonlocal interactions~\cite{STA2}. In the case of atomic two- and three-level systems, the corresponding auxiliary Hamiltonian would include auxiliary laser or microwave interactions~\cite{STA3}. In these systems, the physical implementation of the reference Hamiltonian and auxiliary Hamiltonian differs. However, for a particle with spin in a time-dependent magnetic field, the auxiliary Hamiltonian remains a time-dependent magnetic field~\cite{STA}.
Therefore, we may decompose the given  Hamiltonian
in the BD experiment into a reference Hamiltonian $\hat{H}(t)$ that defines the evolution path of the spin state  and a corresponding auxiliary Hamiltonian $\hat{H}_{\textrm{aux}}(t)$. This allows the given Hamiltonian to precisely track the defined spin state and naturally accumulate dynamical and topological phase.
Then we  rewrite the  magnetic field $\textbf{B}_{0}(t)$ as
\begin{eqnarray}
\label{cichangxiuzheng}
\textbf{B}_{0}(t)=B\textbf{S}(t)-\frac{1}{\kappa}\textbf{S}(t)\times \partial_{t}\textbf{S}(t),
\end{eqnarray}
the first term in Eq.~(\ref{cichangxiuzheng}) corresponds to the reference Hamiltonian $\hat{H}(t)=-\kappa B\textbf{S}(t)\cdot \hat{S}$,
and the second term corresponds to the auxiliary Hamiltonian
$\hat{H}_{\textrm{aux}}(t)=\textbf{S}(t)\times \partial_{t}\textbf{S}(t) \cdot \hat{S}$.
The corresponding evolution $\textbf{S}(t)$ is similar to driving a spin to undergo uniform rotation on a cone with an opening angle of $2\theta$.
\begin{eqnarray}
\begin{aligned}
\textbf{S}(t)=\sin\theta(\cos\frac{2\pi v t}{L} \textbf{e}_{x}+\sin\frac{2\pi v t}{L}\textbf{e}_{y})+\cos\theta\textbf{e}_{z}.
\end{aligned}
\end{eqnarray}
Further, one can obtain
\begin{eqnarray}
\begin{aligned}
\textbf{B}_{0}(t)=&(B+\frac{2\pi v}{\kappa L}\cos\theta)\sin\theta\cos\frac{2\pi vt}{L} \textbf{e}_{x}\\
&+(B+\frac{2\pi v}{\kappa L}\cos\theta)\sin\theta\sin\frac{2\pi v t}{L}\textbf{e}_{y}\\
&+(B\cos\theta -\frac{2\pi v}{\kappa L}\sin^{2}\theta)\textbf{e}_{z},
\end{aligned}
\end{eqnarray}
the corresponding solution is
\begin{eqnarray}
\label{xiuzheng}
\begin{aligned}
B&=\frac{B_{0}(B_{0}+\frac{2\pi v \cos\phi}{\kappa L})}{\sqrt{B_{0}^{2}+\frac{4\pi^{2}v^{2}}{\kappa^{2}L^{2}}+\frac{4 B_{0}\pi v\cos\phi}{\kappa L}}},\\
\theta&=\arctan[\frac{B_{0}\sin\phi}{B_{0}\cos\phi+\frac{2\pi v}{\kappa L}}].
\end{aligned}
\end{eqnarray}

When the magnetic field $\textbf{B}_{0}(t)$ undergoes a nonadiabatic cyclic evolution by the closed curve $\textbf{C}$ as seen from the origin $\textbf{B}_{0}=0$, at which point each substate $|s_{n}\rangle$ acquires a dynamical phase factor of $\textrm{exp}(i s_{n}\varepsilon)$ with $\varepsilon=\kappa\int_{0}^{T}B dt$~\cite{BD}, as well as a topological phase factor of $\textrm{exp}(i s_{n}\gamma)$ with $\gamma=-\Omega$~\cite{BD}.
Here $\hat{\textbf{S}}(t)=\Sigma s_{n}|s_{n}(t)\rangle \langle s_{n}(t)|\,(n=1,2)$~\cite{biaoshu} and $\Omega$ is the spin cone angle $\theta$  subtended by a closed curve in the parameter space. The time $T=L/v$ is for a neutron to travel along the z-axis in the helical magnetic field.

When the polarized neutrons are prepared, their polarization can be viewed as being divided into two substates aligned either parallel or antiparallel to the direction of the magnetic field. During their interaction with the magnetic field, these two substates accumulate different phase shifts. The phase shift can be detected through the coherent superposition of these two substates. Upon their superposition at the end of the field region, it results in an overall spin rotation. This polarized neutron interferometry technique can be employed to manifest the  topological phase in neutron spin rotation~\cite{BD,BD2,BD3,BD4,BD5,BD6,BD6}.
Here the exact wave function of the neutron at time $t=T$ is
\begin{eqnarray*}
\begin{aligned}
|\Psi(t,L)\rangle&=\cos\frac{\theta}{2}\textrm{exp}\left[\frac{i}{2}(\kappa BT-2\pi(1-\cos\theta))\right]|s_{1}(t)\rangle\\
&+\sin\frac{\theta}{2}\textrm{exp}\left[-\frac{i}{2}(\kappa BT-2\pi(1-\cos\theta))\right]|s_{2}(t)\rangle\\
&\equiv a(T)|\uparrow\,\rangle+b(T)|\downarrow\,\rangle,
\end{aligned}
\end{eqnarray*}
for the neutron beam initially in the state $|\uparrow\,\rangle$, which gives the polarization of neutron along the z axis:
\begin{eqnarray}
\label{jihua}
\begin{aligned}
P_{z}&=|a(T)|^{2}-|b(T)|^{2}\\
&=1-2\sin^{2}\theta\sin^{2}\left[\frac{\kappa BT-2\pi(1-\cos\theta)}{2}\right].
\end{aligned}
\end{eqnarray}
When the adiabatic conditions are fulfilled, i.e., $\textbf{B}_{0}(t)$ is both homogeneous and sufficiently strong, and the velocity $v$ of the neutron along the $z$ axis
is small enough, $\theta\simeq\phi$ from Eq.~(\ref{xiuzheng}).
In this case, the corresponding polarization of the neutron of Eq.~(\ref{jihua}) would reduce to $P^{ad}_{z} = 1-2\sin^{2}\phi\sin^{2}\left[\frac{\kappa B_{0}T-2\pi(1-\cos\phi)}{2}\right]$, which is the original result obtained in the BD experiment under adiabatic conditions~\cite{BD,Sun}.

\textit{Discussion and outlook}.\textemdash In this study, during the reexamination the topological phase in the BD experiment, we clarify the previous assumptions that adiabatic conditions are disrupted when neutrons fly at fast velocities in a helical magnetic field that is neither homogeneous nor strong enough, leading to the destruction of the topological phase.
Even when the adiabatic conditions are disrupted in the BD experiment,  the neutron spin rotation can still acquire a nonadiabatic topological phase. This is because  the nonadiabatic topological phase in neutron spin rotation reflects the properties of the entire system, encompassing both the spatial configuration of the magnetic field and the motion state of the neutron. This indicates that under nonadiabatic conditions, the manifested nonadiabatic topological phase not only rely on their geometric history but also depend on the internal dynamics of the system.  When expressed as a line integral over the loop in the parameter space, these nonadiabatic topological phase become sensitive to the exact rate of change along the loop.

The topological phase is also regarded as the direct observable effect of Aharonov-Bohm phase of the induced gauge potentials in the laboratory frame of reference~\cite{Sun,AB}.
Here we provide the exact solution for the polarization of neutrons flying along a more general static helical magnetic field in the moving frame of reference. Consequently, the same exact solution for the polarization of neutrons can also be obtained in the laboratory frame of reference. In the laboratory frame, the polarization of the neutron is determined by the induced gauge potential, which comprises effective vector and scalar potentials. The induced gauge potential originates from the adiabatic variable separation of the spin and spatial coordinates of a particle in an external field, according to the generalized Born-Oppenheimer (BO) approximation.
Therefore, Eq.~(\ref{jihua}) would guide us on how to extend the generation of the induced gauge potential when the BO approximation is invalid~\cite{BO,BO2,Sun2}. This offers a promising method for optimizing the generation of artificial gauge potentials for neutral atoms~\cite{other2,Li}.

\begin{figure}[htbp]
\begin{center}
\includegraphics[width=250pt]{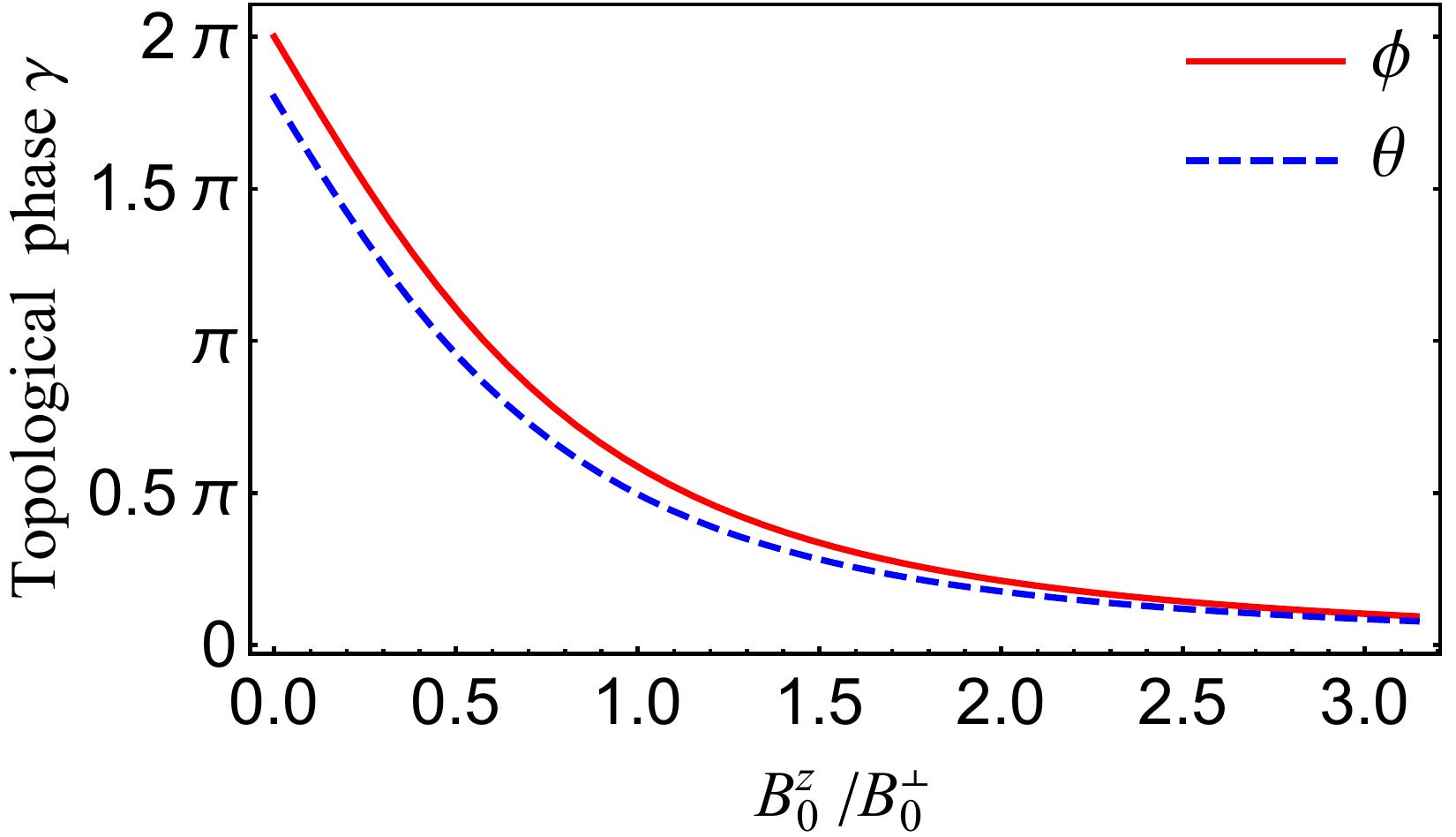}
\caption{Topological phase $\gamma$ at different solid angles  of the helical magnetic field $\textbf{B}_{0}(t)$, for experimental data see Ref.~\cite{BD}.}
\label{fig2}
\end{center}
\end{figure}

We derived the specific expression for the  topological phase in neutron spin rotation without the adiabatic regime in general cases.
As illustrated in Fig.~\ref{fig2}, we perform numerical simulations to  plot the  topological phase based on the spin cone angle $\theta$ (blue dashed line) and the static helical magnetic field cone angle $\phi$ (red solid line).
The numerical values of  the nonadiabatic topological phase derived from the spin cone angle demonstrate a strong concurrence with the experimental data obtained from the BD experiment~\cite{BD}.
In this case, the Berry phase (red solid line) represents the first-order approximation of the  topological phase in neutron spin rotation under adiabatic conditions.
Here, we present a more direct quantification criterion for adiabatic conditions.

This study presents a more comprehensive exploration of the topological phase in neutron spin rotation, demonstrating its concurrence with experimental data from the BD experiment under conditions that surpass traditional adiabatic limits. Furthermore, it illuminates a broader range of spin dynamics phenomena in neutron systems. These findings have the potential to advance various applications, including precision measurements, quantum computation, and quantum information processing.

\textit{Conclusion}.\textemdash To summarize, the utilization of time-dependent rotating magnetic fields enables us, extending and building upon the measurements of Berry phase with polarized neutrons in the BD experiment, to unveil topological phase effects even for nonadiabatic evolutions.
The nonadiabatic topological phase discovered in this study have the potential to inspire further investigations into innovative experimental setups and theoretical frameworks, such as how to reasonably extend the generation of the induced gauge potential when the BO approximation is invalid.

I acknowledge  discussions with Nan Zhao and Hua Zheng. This work was supported by  the National Natural Science Foundation of China (Grants
No. U223040003, and No. 12088101) .

\end{document}